# A Multivariate Equivalence Test Based on Mahalanobis Distance with a Data-Driven Margin

Chao Wang[1], Yu-Ting Weng, Shaobo Liu, Tengfei Li, Meiyu Shen, and Yi Tsong

Office of Biostatistics, Center for Drug Evaluation and Research, U.S. Food and Drug Administration, Silver Spring, Maryland

Abstract

Multivariate equivalence testing is needed in a variety of scenarios for drug development. For example, drug products obtained from natural sources may contain many components for which the individual effects and/or their interactions on clinical efficacy and safety cannot be completely characterized. Such lack of sufficient characterization poses a challenge for both generic drug developers to demonstrate and regulatory authorities to determine the sameness of a proposed generic product to its reference product. Another case is to ensure batch-to-batch consistency of naturally derived products containing a vast number of components, such as botanical products. The equivalence or sameness between products containing many components that cannot be individually evaluated needs to be studied in a holistic manner. Multivariate equivalence test based on Mahalanobis distance may be suitable to evaluate many variables holistically. Existing studies based on such method assumed either a predetermined constant margin, for which a consensus is difficult to achieve, or a margin derived from the data, where, however, the randomness is ignored during the testing. In this study, we propose a multivariate equivalence test based on Mahalanobis distance with a data-drive margin with the randomness in the margin considered. Several possible implementations are compared with existing approaches via extensive simulation studies.

[1] Corresponding author. Email: Chao.Wang@fda.hhs.gov. Tel: +1 (301) 796-1407. Address: U.S. Food and Drug Administration, 10903 New Hampshire Ave., Silver Spring, MD 20903.

# 1   Introduction

Multivariate equivalence testing is needed in a variety of scenarios for drug development. For example, drug products obtained from natural sources may contain many components for which the individual effects and/or their interactions on clinical efficacy and safety cannot be completely characterized. Such lack of sufficient characterization poses a challenge for both generic drug developers to demonstrate and regulatory authorities to determine the sameness of a proposed generic product to its reference product. Another case is to ensure batch-to-batch consistency of naturally derived products containing a vast number of components, such as botanical products. The equivalence or sameness between products containing many components that cannot be individually evaluated needs to be studied in a holistic manner.

The equivalence of an endpoint for two products is frequently assessed via their mean values. In this paper, we assume the parameter of interest to be a $p$-dimensional real-valued vectors, $X_T$ for a test product (T) and $X_R$ for a reference product (R), and the aim is to establish the sameness or equivalence between $X_T$ and $X_R$.

Assume that $X_I$ follows a multivariate normal distribution with mean $\mu_I$ and covariance matrix $\Sigma_I$, $I = T, R$. For this paper, assume that $\Sigma_T = \Sigma_R = \Sigma$ for some positive-definite matrix $\Sigma$. One metric to assess the difference between $X_T$ and $X_R$ is the squared Mahalanobis distance, $\mathcal{D}_M^2 := \mathcal{D}_M^2(\mu_T, \mu_R, \Sigma) = (\mu_T - \mu_R)'\Sigma^{-1}(\mu_T - \mu_R)$.

Let $X_{I,1}, \ldots, X_{I,n_I}$ be an independent and identically distributed sample for $X_I$, $I = T, R$. Then $\mu_T, \mu_R$, and $\Sigma$ can be estimated respectively by their sample estimates:

$$\hat{\mu}_I = \frac{1}{n_I}\sum_{i=1}^{n_I} X_I$$

$$\hat{\Sigma}_I = \frac{1}{n_I - 1}\sum_{i=1}^{n_I}(X_I - \hat{\mu}_I)'(X_I - \hat{\mu}_I)$$

$$\hat{\Sigma} = \frac{1}{n_T + n_R - 2}\Big((n_T - 1)\hat{\Sigma}_T + (n_R - 1)\hat{\Sigma}_R\Big).$$

An estimate for $\mathcal{D}_M^2(\mu_T, \mu_R, \Sigma)$ may be given by the plug-in estimator $\widehat{\mathcal{D}}_M^2 := \mathcal{D}_M^2(\hat{\mu}_T, \hat{\mu}_R, \hat{\Sigma})$. By Anderson (2003, 176, Theorem 5.2.2), we have

$$\mathcal{T}^2 := \frac{n_T + n_R - p - 1}{(n_T + n_R - 2)p}\frac{n_T n_R}{(n_T + n_R)}\widehat{\mathcal{D}}_M^2 \sim F_{p,n_T+n_R-p-1}\left(\frac{n_T n_R}{n_T + n_R}\mathcal{D}_M^2\right). \quad (1.1)$$

Note that the non-central parameter for the $F$ distribution depends on the unknown parameter $\mathcal{D}_M^2$.

Equivalence tests based on $\mathcal{D}_M^2$ have been proposed in the literature and are generally set up to test the following hypotheses:

$$H_0: \mathcal{D}_M^2 \geq \Delta^2 \text{ versus } H_a: \mathcal{D}_M^2 < \Delta^2 \quad (1.2),$$

for some equivalence margin $\Delta^2$.

These equivalence tests vary on how $\Delta^2$ is defined and how the testing procedure is implemented. For instance, Wellek (2010, sec. 8) considered the case that $\Delta^2$ is a pre-specified constant, in which the decision rule of an exact level-$\alpha$ test can be given by rejecting $H_0$ if $\mathcal{T}^2 < F_{p,n_T+n_R-p-1;\alpha}\left(\frac{n_T n_R}{n_T+n_R}\Delta^2\right)$, where $F_{d_1,d_2,c}(\alpha)$ denotes the lower $\alpha$-quantile of the non-central F distribution with $d_1$ and $d_2$ degrees of freedom and noncentral parameter $c$. However, a consensus on the value of the constant $\Delta^2$ which can be applied to a wide range of problems is very difficult to achieve.

Motived by the "global similarity limit" by Tsong et al. (1996), Hoffelder (2019) proposed a random equivalence margin for dissolution profile comparison

$$\hat{\Delta}^2 := D'\hat{\Sigma}^{-1}D, \qquad (1.3)$$

where $D$ is a predefined constant vector, e.g, $D = (10, \cdots, 10)'$.

Hoffelder (2019) called his approach T2EQ and proposed the following decision rule:

$$H_0 \text{ is rejected if } \mathcal{T}^2 < F_{p,n_T+n_R-p-1,\frac{n_T n_R}{n_T+n_R}\hat{\Delta}^2}(\alpha). \qquad (1.4)$$

In addition, Hoffelder (2019) also considered bootstrapping the Mahalanobis distance. He stated that a small simulation study was performed but not reported with percentile and BCa method (Efron and Tibshirani 1993) as well as the basic bootstrap according to Davison and Hinkley (1997).

The T2EQ approach used a margin that depends on the data. However, the decision rule did not take the randomness in the margin into account. Some of the empirical type I error rates reported in Hoffelder (2019) were significantly less than the nominal type I error rate. We suspect that this is due to the random margin that is plugged in as the noncentral parameter of an F distribution. It is generally more preferable that such randomness is taken into account in the decision rule. In this paper we proposed an approach with the randomness accounted for to improvement the control of type I error rate.

The paper is organized as follows. Section describes the methods. Section presents simulation studies to compare various methods. Section summarizes the paper.

## 2   Method

In this paper, we propose a margin that is similar to the random margin in Eq (1.3) and consider the fact that the margin must be estimated in the test. The proposed margin has the following form,

$$\Delta^2 := D'\Sigma^{-1}D, \qquad (2.1)$$

where $D \in \mathbb{R}^p$ is a constant vector, independent of the data for comparison, $\Sigma$ is the true common variance-covariance matrix for both products.

Note that the hypotheses in Eq (1.2) are equivalent to

$$H_0: \mathcal{D}_M^2(\mu_T, \mu_R, \Sigma_C) - \Delta^2 \geq 0 \quad (2.2)$$
$$H_1: \mathcal{D}_M^2(\mu_T, \mu_R, \Sigma_C) - \Delta^2 < 0 \quad (2.3)$$

Let $\mathcal{A} = \mathcal{D}_M^2(\mu_T, \mu_R, \Sigma_C) - \Delta^2$ be the parameter of interest and

$$\hat{\mathcal{A}} = \hat{\mathcal{D}}_M^2 - \hat{\Delta}^2 \quad (2.4)$$

be the sample estimate of $\mathcal{A}$.

The sample estimates $\hat{\mathcal{D}}_M^2$ and $\hat{\Delta}^2$ are biased. Anderson (2003, 228, Eq (4)) showed that

$$E[\hat{\mathcal{D}}_M^2] = \frac{n-2}{n-p-3}\left(\mathcal{D}_M^2 + p\left(\frac{1}{n_T} + \frac{1}{n_R}\right)\right)$$

where $n = n_T + n_R$.

It is clear that the bias is always positive and is greater if $p$ is large or $n_T$ and $n_R$ is small. Thus, an unbiased estimate of $\mathcal{D}_M^2$ can be constructed as

$$\hat{\mathcal{D}}_{M,bc}^2 = \frac{n-p-3}{n-2}\hat{\mathcal{D}}_M^2 - p\left(\frac{1}{n_T} + \frac{1}{n_R}\right)$$

In addition, the inverse of a sample variance-covariance matrix is a biased estimate of the precision matrix, i.e., the inverse of the population variance-covariance matrix (see Anderson (2003, 274), page 255, Thm. 7.2.2 for the density of Whishart distribution, Corollary 7.2.3. for the distribution of the sample variance matrix, Page 243, Lamma 7.7.1. for the expectation of the sample variance matrix). Based on the referenced results, the plugged-in estimate of the margin $\hat{\Delta}^2$ is a biased estimate of the proposed margin, $\Delta^2$, and it can be adjusted to be unbiased,

$$\hat{\Delta}_{bc}^2 = \frac{n-p-3}{n-2}D'\hat{\Sigma}^{-1}D.$$

We propose to use the following test statistics based on the unbiased estimate of the population difference

$$\hat{\mathcal{A}}_{bc} = \hat{\mathcal{D}}_{M,bc}^2 - \hat{\Delta}_{bc}^2$$
$$= \frac{n-p-3}{n-2}\hat{\mathcal{A}} - p\left(\frac{1}{n_T} + \frac{1}{n_R}\right) \quad (2.5)$$

Because the sampling distributions of $\hat{\mathcal{D}}_M^2$ and $\hat{\Delta}^2$ and thus $\hat{\mathcal{A}}_{bc}$ depend on the unknown $\mathcal{D}$ and $\Sigma$, the test procedure is implemented via bootstrap (Efron and Tibshirani 1993).

The percentile bootstrap procedure with test statistic $\hat{\mathcal{A}}_{bc}$ is given below:

- Let $\left(X_{T,i}^{*,b}\right)_{i=1}^{n_T}$ and $\left(X_{R,i}^{*,b}\right)_{i=1}^{n_R}$ denote the $b$-th bootstrap sample for $X_T$ and $X_R$ by sampling with replacement from each group of $(X_{T,i})_{i=1}^{n_T}$ and $(X_{R,i})_{i=1}^{n_R}$ independently respectively for $b = 1,\ldots,B$, where $B$ denotes the number of bootstrap samples.
- Let $\mathcal{A}_{bc,(b)}^{*}$ be the sample estimate for $\hat{\mathcal{A}}_{bc}$ calculated with the $b$-th bootstrap sample.
- Let $\mathcal{A}_{bc,1-\alpha}^{*}$ be the $100 \times (1-\alpha)$ percentile of the $B$ Bootstrap replications $\left(\mathcal{A}_{(b)}^{2,*}\right)_{b=1}^{B}$.
- Reject $H_0$ if $\mathcal{A}_{bc,1-\alpha}^{*} < 0$.

It was conjectured that the above percentile bootstrap may not have satisfactory performance in controlling type I error. More sophisticated bootstrap methods such as BCa and ABC bootstrap were also considered.

In general, the BCa bootstrap procedure in Efron and Tibshirani (1993, 185, Equation 14.) was followed. The decision rule for the BCa bootstrap is as follows:

$$H_0 \text{ is rejected if } \mathcal{A}_{bc,f(1-\alpha)}^{*} < 0, \qquad (2.6)$$

where $f(1-\alpha)$ is the BCa-adjusted value of $1-\alpha$,

$$f(1-\alpha) = \Phi\left(\hat{z}_0 + \frac{\hat{z}_0 + z^{(1-\alpha)}}{1 - \hat{a}(\hat{z}_0 + z^{(1-\alpha)})}\right). \qquad (2.7)$$

The term $\Phi$ in Eq (2.7) is the CDF of the standard normal distribution, $\hat{z}_0$ is the bias-correction term and $\hat{a}$ is the acceleration term.

The $\hat{z}_0$ is given by Efron and Tibshirani (1993, 186, Equation 14.14)

$$\hat{z}_0 = \Phi^{-1}\left(\frac{\#\{\mathcal{A}_{bc,(b)}^{*} < \hat{\mathcal{A}}_{bc}\}}{B}\right)$$

The acceleration term $\hat{a}$ can be estimated by a jackknife procedure in Efron and Tibshirani (1993, 186, Equation 14.15).

The BCa bootstrap turned out to perform not better than the percentile bootstrap in both scenarios considered in the simulation study reported below. It is possible to improve the performance of both BCa and percentile bootstrap via bootstrap calibration, but the computational cost is too large for an accurate simulation due to the double-layer bootstrap loop. In the following, we consider an ABC bootstrap technique. Because ABC bootstrap is usually applied to a one-sample problem, we first transform the two-sample comparison into a one sample problem using the technique in Anderson (2003, 188).

To explore the performance, We considered the case when $n_T = n_R = n_c$. Since the $(X_{T,i})$ and $(X_{R,i})$ are assumed independent samples, we can define $X_i = X_{T,i} - X_{R,i}$ be the "paired" difference. Then

$$X_i \underset{IID}{\sim} N(\mu_T - \mu_R, 2\Sigma),$$

Then

$$\bar{X} = \frac{1}{n_c}\sum_{i=1}^{n_c} X_i \sim N\left(\mu_T - \mu_R, \frac{2}{n_c}\Sigma_c\right),$$

and

$$S_c := \frac{1}{2}\frac{1}{n_c - 1}\sum_{i}^{n_c}(X_i - \bar{X})(X_i - \bar{X})'$$

where $2(n_c - 1)S_c = \sum_{i}^{n_c}(X_i - \bar{X})(X_i - \bar{X})' =_d \sum_{i=1}^{n_c-1} Z_i Z_i'$ where

$$Z_i \underset{IID}{\sim} N(0, 2\Sigma_c)$$

The ABC bootstrap can be calculated based on the test statistic:

$$\bar{X}'S_c^{-1}\bar{X} - D'S_c^{-1}D \qquad (2.8)$$

and implemented by abcnon.

Then a calibration procedure by Efron and Tibshirani (1993) can be performed based on ABC bootstrap.

## 3   Simulation studies

To evaluate the performance of the proposed tests, we performed extensive simulation studies. We focused on two scenarios as in Hoffelder (2019). The aim was to compare the testing procedures that are listed below:

- (T2EQ): The T2EQ approach in Hoffelder (2019), with the testing procedure summarized in Eq. (1.4).

- (T2EQTM): The T2EQ approach with the non-central parameter given by the true value. Note that this is only feasible in simulation studies where the true noncentral parameter is known. It can offer insight regarding the impact of plugging in sample covariance matrix to the non-central parameter on the performance of the T2EQ approach and the variability in the simulation.

- (PctBootstrapOnMD): The percentile bootstrap implementation of the T2EQ approach.

- (PctBootstrapOnDif): The percentile bootstrap implementation based on the test statistics in Eq. (2.4).

- (PctBootstrapOnDifBC): The percentile bootstrap implementation based on the test statistic in Eq. (2.5).

- (BCaBootstrapOnMD): The BCa bootstrap implementation of the T2EQ approach.

- (BCaBootstrapOnDif): The BCa bootstrap implementation based on the test statistics in Eq. (2.4).

- (BCaBootstrapOnDifBC): The BCa bootstrap implementation based on the test statistics $\hat{A}_{bc}$ in Eq. (2.5).

- (CalibABC): The bootstrap calibration of the ABC bootstrap on test statistics in Eq. (2.8) for the paired difference.

Both the empirical type I error rate and power were assessed. The parameters used in the simulation were largely the same as in Hoffelder (2019) and designed to cover both 3-dim and 4-dim comparisons. In addition, the underlying data distribution can be multivariate normal or log-normal distribution.

The parameter sets are given by the two groups below.

- Group 1: $\mu_1 = (20, 70, 88)'$, $D = (10, 10, 10)'$, $E = (8, 8, 8)'$,

$$\Sigma = \begin{bmatrix} 120 & 39 & -9 \\ 39 & 146 & 111 \\ -9 & 111 & 113 \end{bmatrix}$$

- Group 2: $\mu_1 = (31, 61, 83, 94)'$, $D = (10, 10, 10, 10)'$, $E = (8, 8, 8, 8)'$,

$$\Sigma = \begin{bmatrix} 32 & 52 & 38 & 26 \\ 52 & 87 & 66 & 44 \\ 38 & 66 & 65 & 35 \\ 26 & 44 & 35 & 30 \end{bmatrix}$$

For each group, $\mu_2 = \mu_1 + D$ for type I error study, and $\mu_2 = \mu_1 + E$ for power study with the respective $\mu_1$, $D$, and $E$ parameters.

The simulation studies with the following scenarios were performed:

- Under normal distribution:
    - S1 (empirical type I error in $\mathbb{R}^3$): With parameters in Group 1, $\mu_2 = \mu_1 + D$, $X_R \sim N(\mu_1, \Sigma)$, $X_T \sim N(\mu_2, \Sigma)$,
    - S2 (empirical type I error in $\mathbb{R}^4$): With parameters in Group 2, $\mu_2 = \mu_1 + D$, $X_R \sim N(\mu_1, \Sigma)$, $X_T \sim N(\mu_2, \Sigma)$,
    - S3 (empirical power in $\mathbb{R}^3$): With parameters in Group 1, $\mu_2 = \mu_1 + E$, $X_R \sim N(\mu_1, \Sigma)$, $X_T \sim N(\mu_2, \Sigma)$,
    - S4 (empirical power in $\mathbb{R}^4$): With parameters in Group 2, $\mu_2 = \mu_1 + E$, $X_R \sim N(\mu_1, \Sigma)$, $X_T \sim N(\mu_2, \Sigma)$.
- Under log-normal (LN) distribution:

- S5 (empirical type I error in $\mathbb{R}^3$): With parameters in Group 1, $\mu_2 = \mu_1 + D$, $X_R \sim LN(\mu_1, \Sigma), X_T \sim LN(\mu_2, \Sigma)$,
- S6 (empirical type I error in $\mathbb{R}^4$): With parameters in Group 2, $\mu_2 = \mu_1 + D$, $X_R \sim LN(\mu_1, \Sigma), X_T \sim LN(\mu_2, \Sigma)$,
- S7 (empirical power in $\mathbb{R}^3$): With parameters in Group 1, $\mu_2 = \mu_1 + E$, $X_R \sim LN(\mu_1, \Sigma), X_T \sim LN(\mu_2, \Sigma)$,
- S8 (empirical power in $\mathbb{R}^4$): With parameters in Group 2, $\mu_2 = \mu_1 + E$, $X_R \sim LN(\mu_1, \Sigma), X_T \sim LN(\mu_2, \Sigma)$.

Given the true mean vectors and covariance matrices, the data for test and reference products were simulated from the corresponding multivariate normal or log-normal distribution, with equal sample size for test and reference products being 12, 24, 36, 48, 96, or 120. Each sample was tested by all testing procedures aforementioned. Empirical type I error rates and powers were based on $3.5 \times 10^5$ replications so that the standard errors of the empirical rates are about 0.01.

The results are illustrated in Fig. 3.1 to 3.8 for the 8 scenarios S1 to S8, respectively. The results largely agreed with those tests reported by Hoffelder (2019). Fig. 3.1 to 3.4 show the improvement of bootstrap-type procedures with the test statistics constructed on the difference over T2EQ. Although many bootstrap-type procedures were compared, within all test procedures based on the difference test statistics, no single procedure outperformed all the others. Table 3.1 further summarizes the mean absolute difference in percentage from the nominal size of 5% for the scenarios S1, S2, S5, and S6. On average the PctBootstrapOnDifBC method had the least deviation from the nominal size. While CaliABCBootstrapOnPairedDif was expected to show the best performance and it had the best performance for S1 and S5, the performance deteriorated for the cases in S2 and S6 with small sample sizes, which might be due to numerical issues caused by the small sample size and relatively high dimension, and thus did not have the best overall performance for the scenarios considered.

The empirical power curves were comparable to those reported in Hoffelder (2019) as well. The different test procedures showed different power curves for the scenarios considered and they were comparable when the distribution was changed from normal distribution to log-normal distribution.

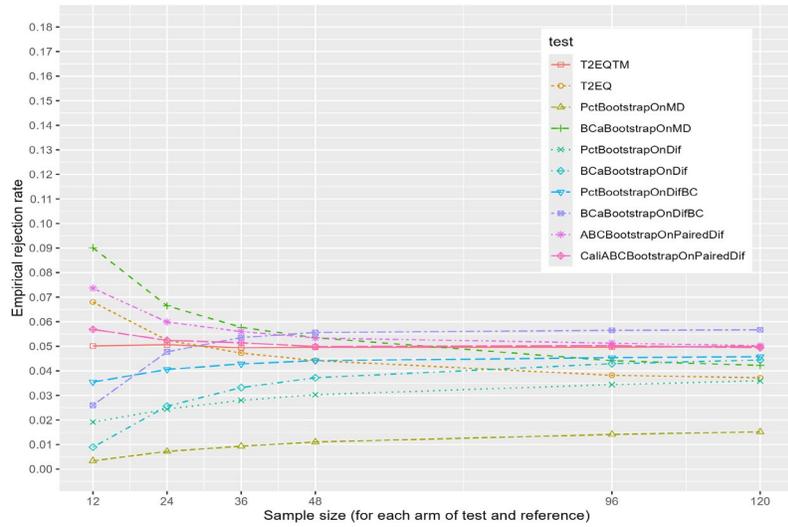

*Figure 3.1: Empirical type I error for Scenario 1.*

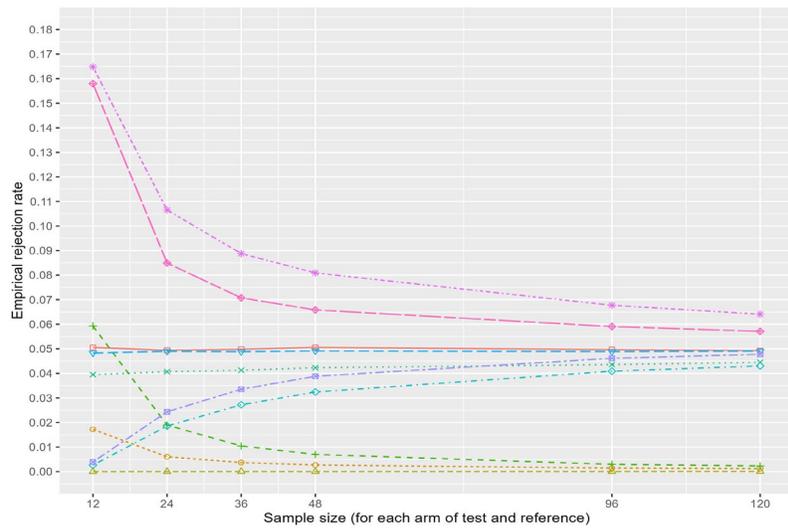

*Figure 3.2: Empirical type I error for Scenario 2.*

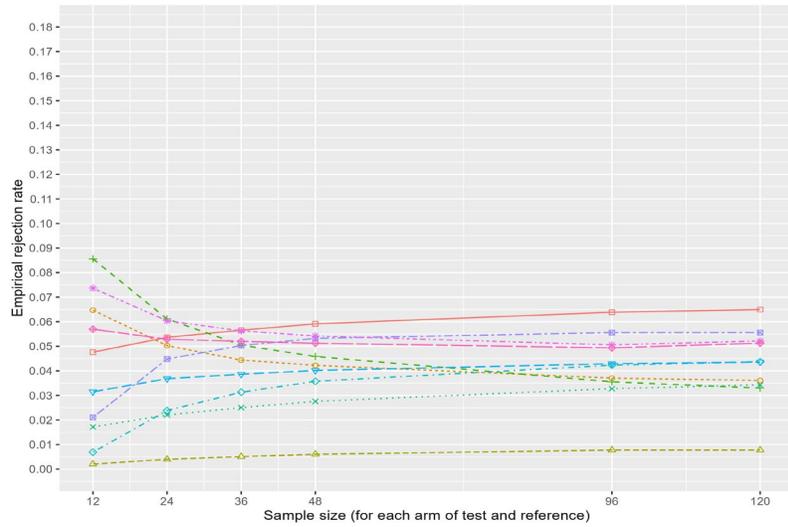

*Figure 3.3: Empirical type I error for Scenario 5.*

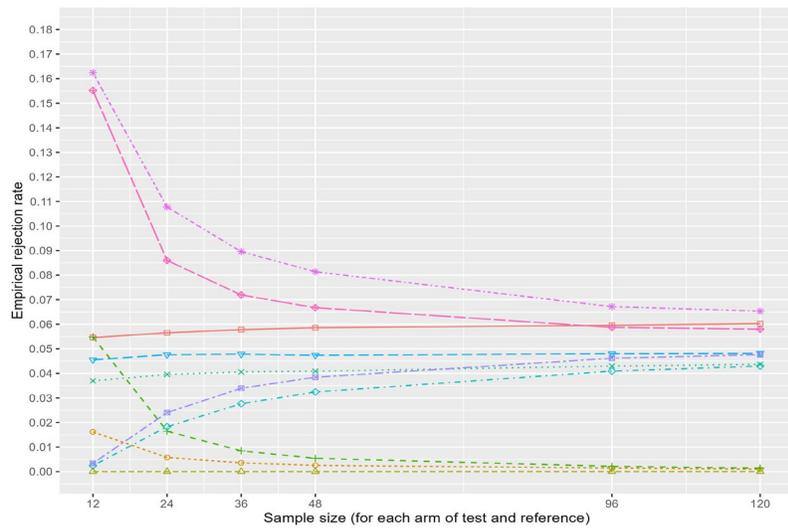

*Figure 3.4: Empirical type I error for Scenario 6.*

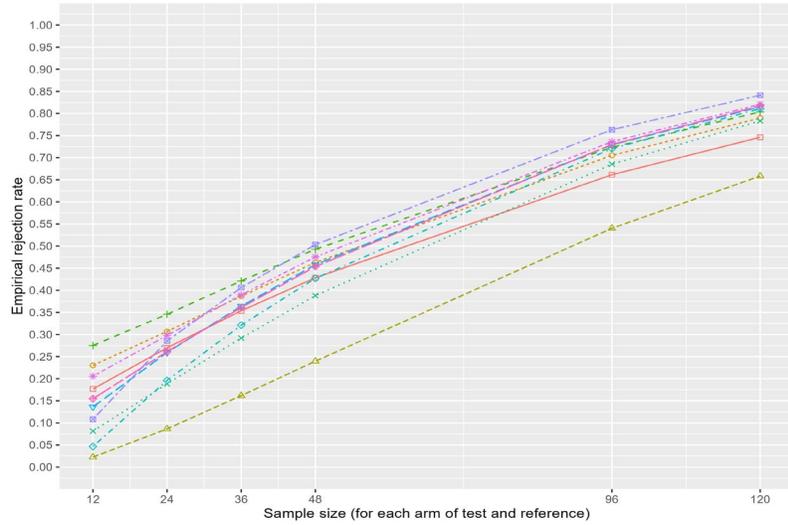

*Figure 3.5: Empirical power for Scenario 3, c.f. Table 3 in Hoffedler (2019).*

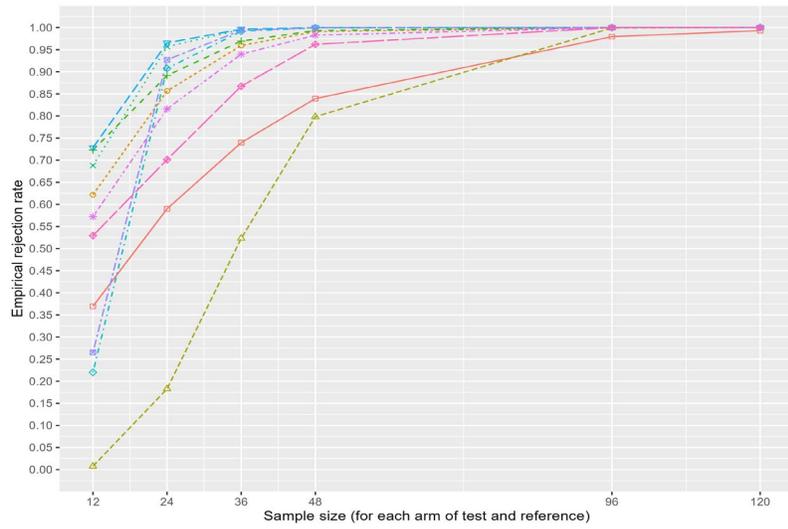

*Figure 3.6: Empirical power for Scenario 4, c.f. Table 3 in Hoffedler (2019).*

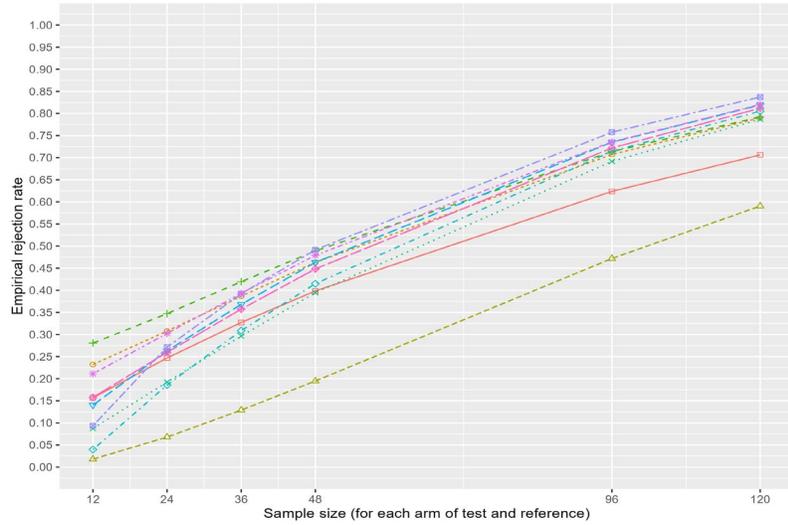

*Figure 3.7: Empirical power for Scenario 7, c.f. Table 3 in Hoffedler (2019).*

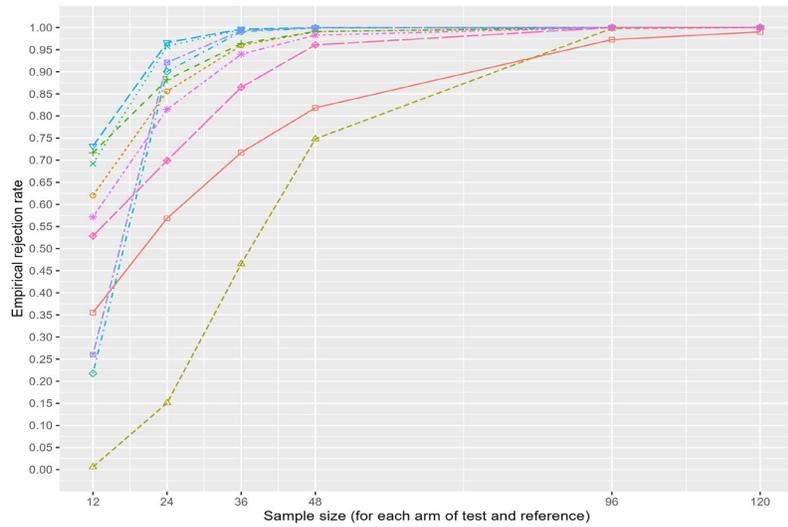

*Figure 3.8: Empirical power for Scenario 8, c.f. Table 3 in Hoffedler (2019).*

|                            | S1   | S2   | S5   | S6   | Average |
|---------------------------:|------|------|------|------|---------|
| T2EQTM                     | 0.03 | 0.05 | 0.84 | 0.79 | 0.43    |
| PctBootstrapOnDifBC        | 0.77 | 0.11 | 1.11 | 0.26 | 0.56    |
| BCaBootstrapOnDifBC        | 0.81 | 1.76 | 0.81 | 1.77 | 1.29    |
| PctBootstrapOnDif          | 2.13 | 0.80 | 2.35 | 0.92 | 1.55    |
| CaliABCBootstrapOnPairedDif| 0.20 | 3.26 | 0.25 | 3.28 | 1.75    |
| BCaBootstrapOnDif          | 1.80 | 2.25 | 1.94 | 2.26 | 2.06    |
| BCaBootstrapOnMD           | 1.36 | 3.63 | 1.38 | 3.68 | 2.51    |
| ABCBootstrapOnPairedDif    | 0.74 | 4.55 | 0.79 | 4.56 | 2.66    |
| T2EQ                       | 0.90 | 4.46 | 0.92 | 4.49 | 2.69    |
| PctBootstrapOnMD           | 4.00 | 5.00 | 4.46 | 5.00 | 4.61    |

*Table 3.2 Summary of mean absolute difference (percentage point reported only) from the nominal size 5% for scenarios S1, S2, S5, and S6, and average over all four scenarios. The tests are sorted in ascending order by the average in the last column.*

## 4 Conclusion and Discussion

The T2EQ approach uses a equivalence margin calculated with a pooled sample estimate of the covariance matrix but does not consider the variability in the margin. This leads to inaccurate distribution of the test statistics and contributes to the inadequate control of type I error rate. We proposed to consider the variability in the equivalence margin and constructed a test statistic based on the difference of the squared Mahalanobis distance and the margin. As the theoretical distribution of the test statistic depends on the unknown true parameters, we considered various bootstrap procedures to implement the test. The simulation results showed that the proposed test improves greatly over the T2EQ and different bootstrap procedures result in difference performance. In general, the percent bootstrap based on the bias-corrected test statistics outperforms the other bootstrap procedures considered. More complicated bootstrap procedures, such as, BCa and calibration, improve the percent bootstrap procedures in some cases, but may encounter issues in when the dimension of the data is high compared with the sample size.

The following issues may be addressed in future reserach. The comparison of the various bootstrap procedure cannot conclude which procedure performs the best. Additionally, the dimensionality of the data to be compared can affect the performance and numeric stability of the test procedure and thus one may consider such comparison in a high-dimensional situation.

## References

Anderson, T. W. 2003. *An Introduction to Multivariate Statistical Analysis*. Wiley Series in Probability and Statistics. Hoboken, New Jersey: Wiley.
https://books.google.com/books?id=Cmm9QgAACAAJ.

Davison, Anthony Christopher, and David Victor Hinkley. 1997. *Bootstrap Methods and Their Application*. 1. 32 Ave. of the Americas, New York, NY 10013, USA: Cambridge university press.

Efron, Bradley, and Robert J Tibshirani. 1993. "An Introduction to the Bootstrap." *Monographs on Statistics and Applied Probability* 57: 1–436.

Hoffelder, Thomas. 2019. "Equivalence Analyses of Dissolution Profiles with the Mahalanobis Distance." *Biometrical Journal* 61 (5): 1120–37.

Tsong, Yi, Thomas Hammerstrom, Pradeep Sathe, and Vinod P Shah. 1996. "Statistical Assessment of Mean Differences Between Two Dissolution Data Sets." *Drug Information Journal* 30 (4): 1105–12.

Wellek, Stefan. 2010. *Testing Statistical Hypotheses of Equivalence and Noninferiority*. CRC Press.